\documentclass{iopart}

\usepackage{graphics}
\usepackage{graphicx}
\usepackage{epsfig}

\begin{document}

\title{Factorised steady states for multi-species mass transfer models} 
\author{T. Hanney}
\address{SUPA and School of Physics, University of Edinburgh,
Mayfield Road, Edinburgh, EH9 3JZ, UK}
\ead{tom.hanney@ed.ac.uk}


\begin{abstract}
A general class of mass transport models with $Q$ species of conserved
mass is considered. The models are defined on a lattice with parallel
discrete time update rules. For one-dimensional, totally asymmetric
dynamics we derive necessary and sufficient conditions on the mass
transfer dynamics under which the steady state factorises. We
generalise the model to mass transfer on arbitrary lattices and
present sufficient conditions for factorisation. In both cases,
explicit results for random sequential update and continuous time
limits are given.
\end{abstract}

\section{Introduction}
A wide variety of physical phenomena may be characterised by the
stochastic transfer of one or more conserved quantities between
regions of space. Examples include traffic flow \cite{CSS00, OEC98},
shaken granular systems \cite{MWL02}, sandpile dynamics \cite{J05},
cluster aggregation and fragmentation \cite{MKB98} and the dynamics of
phase separation \cite{KLMST02}. However, for models of such
processes, which are devised on the basis of some putative microscopic
dynamics in order to capture the essential elements of these systems,
there exists no general prescription to identify the steady state, if
indeed it exists at all.

Among the models that have been proposed, particularly well studied
examples are the zero-range process (ZRP) \cite{S70, EH05} and the
asymmetric random average process \cite{KG00, RM00}. These models
involve mass transfer between the sites of a lattice. Chief among the
reasons for the widespread interest in these models is that they can
be solved exactly in the steady state, a virtue which has been
exploited to substantially develop our understanding of jamming
transitions and phase separation phenomena. In the framework of
zero-range processes, these phenomena may be understood in terms of
condensation --- the property that a single site may acquire a finite
fraction of the total mass in the system. Moreover, the steady state
assumes a relatively simple form, namely, a factorised form, in which
the steady state is expressed as a product of factors, one factor for
each site of the system.

For models with a single conservation law and mass transfer rules
which depend only on the details of the departure site, the conditions
on the hop rates under which the steady state factorises have now been
established rather completely \cite{EMZ06}. However systems with many
conservation laws are also of wide interest \cite{schutz review}. In
the soluble cases identified so far, the inclusion of additional
conservation laws has been shown to enrich the phase behaviour
dramatically and introduces novel mechanisms of condensation
\cite{HE04}. Practical applications of these mass transfer models with
multiple conservation laws include shaken polydisperse granular
systems \cite{MMWL02} and directed networks \cite{DMS03, AEH}.

It is the purpose of this article to derive conditions on the mass
transfer dynamics under which the steady state assumes a factorised
form for a very wide class of models with an arbitrary number $Q$ of
conservation laws. In section \ref{1d model} we define a
one-dimensional, totally asymmetric model with $Q$ species of
conserved mass and present the factorised steady state. The proof and
the necessary and sufficient conditions under which it holds are given
in section \ref{derivation}. The model is defined with discrete-time
parallel update, which contains the random sequential update and
continuous time limits shown in section \ref{continuous time}. The
generalisation to a $Q$-species mass transfer model on an arbitrary
lattice is presented in section \ref{generalisation}, where sufficient
conditions for factorisation of the steady state are stated. A number
of models which represent applications of the steady states derived
here are provided throughout.

\section{$Q$-species mass transfer model in one dimension}
\label{1d model}
The model is defined on a periodic chain containing $L$ sites. $Q$
species of masses reside on the lattice. Associated with each site $i$
is a mass vector $\underline{m_i}$, the elements of which represent
the mass of each species $q=1, \ldots, Q$ present at $i$:
$\underline{m_i} = (m_i^{(1)}, \ldots, m_i^{(Q)})$. The mass of
species $q$ at site $i$, $m_i^{(q)}$, is a continuous variable.

The dynamics are defined in discrete time, such that the mass vectors
at each site are updated in parallel during each time step, and the
mass is transferred to the nearest neighbour site to the right. The
mass of species $q$ transferred from site $i$ to site $i{+}1$ during
the time step is $\mu_{i}^{(q)}$.  The vector $\underline{\mu}_{i} =
(\mu_{i}^{(1)}, \ldots, \mu_{i}^{(Q)})$ represents the masses of all
species which move from $i$ to $i{+}1$.

The definition of the model is completed by defining the
dynamics. These are specified through the mass transfer probabilities
$\psi_i( \underline{\mu}_{i} | \underline{m}_i )$ which determine the
stochastic variables $\mu_{i}^{(q)}$. Hence, the probability of
transferring the masses $\underline{\mu}_{i}$ from $i$ to $i{+}1$ in a
time step depends on the mass of each species at the departure site,
$\underline{m}_i$, and on the site of departure (through the $i$
subscript on $\psi$). The dynamics conserve to total mass in the
system of each species. Further, the interactions are zero-range ---
they depend only on details of the site of departure of the mass.
 
\subsection{Steady state} \label{ss}
The aim is to establish the conditions under which steady
state of the model defined above assumes a factorised form. Therefore
we look for conditions on the choice of the mass transfer
probabilities under which the steady state weight $F(\{
\underline{m}_i \} )$ to observe the system in a configuration $\{
\underline{m}_i \} = \underline{m}_1, \ldots, \underline{m}_L$ is given by
\begin{equation} \label{fss}
F(\{ \underline{m}_i \} ) = \prod_{i=1}^L f_i( \underline{m}_i )\;,
\end{equation}
i.e., one factor for each site of the system. We find a necessary and
sufficient condition for factorisation is that one can find functions
$v(\underline{\mu})$ and $w_i(\underline{m}-\underline{\mu})$ such that
the mass transfer probabilities can be expressed in the form
\begin{equation} \label{mass transfer probs}
\psi_i( \underline{\mu}_{i} | \underline{m}_i ) = 
\frac{v(\underline{\mu}_{i}) w_i(\underline{m}_i-\underline{\mu}_{i})}
{\left[v * w_i \right](\underline{m}_{i})}\;,
\end{equation}
where the $*$ denotes a convolution (see (\ref{con def})). If the mass
transfer probabilities are of the form (\ref{mass transfer probs}),
then the steady state factorises with single-site weights $f_i(
\underline{m}_i )$ given by
\begin{equation} \label{single-site weights}
f_i( \underline{m}_i ) = \left[v * w_i \right](\underline{m}_{i})\;.
\end{equation}

\section{Proof of the steady state} \label{derivation}
In this section, we turn to the proof of the steady state (\ref{fss})
to (\ref{single-site weights}). The proof follows a similar argument to that
given in \cite{EMZ04, EMZ04a} for the single species case. We write $F(\{
\underline{m}_i \}, t )$ to represent the weight of a configuration at
time $t$. For a single update the evolution of $F(\{ \underline{m}_i
\}, t)$, under the dynamics defined by the mass transfer probabilities,
is described by a master equation which may be written
\begin{eqnarray} \label{master equation}
F(\{ \underline{m}_i' \}, t{+}1 ) &=& \prod_i^L \left[ \int {\rm d}^Q
  \underline{m}_i \int {\rm d}^Q \underline{\mu}_i \psi_i(
  \underline{\mu}_i | \underline{m}_i ) \right. \\ && \qquad
  \times\left. \delta^Q( \underline{m}_i' - \underline{m}_i +
  \underline{\mu}_i - \underline{\mu}_{i{-}1})) \right] F(\{
  \underline{m}_i \}, t ) \;,\nonumber
\end{eqnarray}
where d$^Q \underline{m}_i$ represents an integration over the masses
of all the species at site $i$: d$^Q \underline{m}_i =
\prod_{q=1}^Q{\rm d}m_i^{(q)}$; similarly, the notation
$\delta^Q(\underline{x}) = \prod_{q=1}^Q \delta(x^{(q)})$.

In the steady state, $F(\{ \underline{m}_i \}, t{+}1 ) = F(\{
\underline{m}_i \}, t ) \equiv F(\{ \underline{m}_i \})$. Making these
replacements in (\ref{master equation}) and assuming $F(\{
\underline{m}_i \})$ is given by a factorised form (\ref{fss}) yields 
\begin{eqnarray} \label{factorised ME}
\prod_{i=1}^L f_i( \underline{m}_i' ) &=& \prod_i^L \left[ \int {\rm d}^Q
  \underline{m}_i \int {\rm d}^Q \underline{\mu}_{i}
  \psi_i( \underline{\mu}_{i} | \underline{m}_i )
  \right. \\
&& \qquad \times\left. \delta^Q(\underline{m}_i' - \underline{m}_i
  + \underline{\mu}_i - \underline{\mu}_{i{-}1}) \;f_i(
  \underline{m}_i ) \right] \;.\nonumber 
\end{eqnarray}
The aim of the remainder of this section is to find the conditions
under which the assumption of factorisation holds, i.e., the
conditions under which we can solve (\ref{factorised ME}).

To proceed, we introduce the Laplace transform
\begin{equation}
g_i( \underline{s}_i ) = 
  \int_0^\infty {\rm d}^Q \underline{m}_i \; {\rm e}^{-
  \underline{s}_i \cdot \underline{m}_i} f_i(\underline{m}_i )  \;,
\end{equation}
where $\underline{s}_i = (s_i^{(1)}, \ldots, s_i^{(Q)})$ and
$\underline{s} \cdot \underline{m} = \sum_q s^{(q)} m^{(q)}$. Hence
\begin{eqnarray}
\prod_{i=1}^L g_i( \underline{s}_i ) &=& \prod_i^L \left[ \int {\rm d}^Q
  \underline{m}_i \int {\rm d}^Q \underline{\mu}_{i}
  \psi_i( \underline{\mu}_{i} | \underline{m}_i )
\right. \\ && \qquad \times\left. 
  {\rm e}^{-\underline{s}_i \cdot (\underline{m}_i -\underline{\mu}_i
  +\underline{\mu}_{i{-}1}) } \; f_i(
  \underline{m}_i ) \right] \;.\nonumber 
\end{eqnarray}
At this stage, it is convenient to introduce the function
$\mathcal{P}_i( \underline{\mu}_{i}, \underline{m}_i -
\underline{\mu}_i )$ defined as 
\begin{equation} \label{P}
\mathcal{P}_i( \underline{\mu}_{i}, \underline{m}_i -
  \underline{\mu}_i ) = \psi_i( \underline{\mu}_{i} |
  \underline{m}_i ) \; f_i( \underline{m}_i )\;,
\end{equation}
then, doing the integral over the mass variables for each species,
$m_i^{(q)}$, and re-ordering the sum in the exponential leads to 
\begin{equation}
\prod_{i=1}^L g_i( \underline{s}_i ) = \prod_i^L \left[ \int {\rm d}^Q
  \underline{\mu}_{i}  {\rm e}^{-\underline{s}_i \cdot
  (\underline{m}_i-\underline{\mu}_{i}) - \underline{s}_{i{+}1} \cdot
  \underline{\mu}_i} \; \mathcal{P}_i( \underline{\mu}_{i},
  \underline{m}_i - \underline{\mu}_i ) \right] \;.
\end{equation}
Now, the term in the square brackets on the rhs can be identified with the
function $X_i( \underline{s}_i, \underline{s}_{i{+}1} )$:
\begin{equation} \label{X}
X_i( \underline{s}_i, \underline{s}_{i{+}1} ) = \int {\rm d}^Q
  \underline{\mu}_{i} {\rm e}^{-\underline{s}_i \cdot
  (\underline{m}_i-\underline{\mu}_{i}) - \underline{s}_{i{+}1} \cdot
  \underline{\mu}_i} \; \mathcal{P}_i( \underline{\mu}_{i},
  \underline{m}_i - \underline{\mu}_i )\;,
\end{equation}
hence, we seek a solution to
\begin{equation} \label{gen}
\prod_{i=1}^L g_i( \underline{s}_i ) = \prod_{i=1}^L X_i(
\underline{s}_i, \underline{s}_{i{+}1} )\;,
\end{equation}
in order to establish the conditions for factorisation of the steady
state. Note that we have only rewritten (\ref{factorised ME}) at
this stage --- a necessary and sufficient solution of (\ref{gen})
represents a necessary and sufficient condition for factorisation. 

By taking the logarithm of (\ref{gen}) and differentiating with
respect to $s_i$ then $s_{i{+}1}$, it can be shown \cite{EMZ06, EMZ04}
that a necessary and sufficient condition to solve (\ref{gen}) is
\begin{equation} \label{sufficient solution}
X_i( \underline{s}_i, \underline{s}_{i{+}1} ) = \alpha_{i}
(\underline{s}_i)\, \beta (\underline{s}_{i{+}1})\;,
\end{equation}
where the functions $\alpha_i(\underline{s})$ and
$\beta({\underline{s}})$ are to be determined. This implies
\begin{equation}
g_i(\underline{s}_i) = \alpha_{i}(\underline{s}_i)\,\beta(\underline{s}_i)\;.
\end{equation}
Since this is just a product over Laplace transforms,
its inverse (i.e., $f_i(\underline{m}_i)$) is a convolution  
\begin{equation} \label{con def}
f_i(\underline{m}) = \left[ v * w_i \right] 
(\underline{m}) \equiv \int {\rm d}^Q \underline{\mu} \;
v(\underline{\mu}) \, w_{i}(\underline{m}-\underline{\mu}) \;, 
\end{equation}
as given in (\ref{single-site weights}). 

The functions $\alpha_i(\underline{s})$ and $\beta(\underline{s})$ are
determined by
\begin{equation}
\alpha_{i}(\underline{s}_i) = \int {\rm d}^Q \underline{\nu}\; {\rm
  e}^{-\underline{s}_i \cdot \underline{\nu}} \;
  w_{i}(\underline{\nu})\;,
\end{equation}
where $\underline{\nu} = \underline{m} - \underline{\mu}$, and
\begin{equation}
\beta(\underline{s}_i) = \int {\rm d}^Q \underline{\mu}\; {\rm
  e}^{-\underline{s}_i \cdot \underline{\mu}} \;
  v(\underline{\mu})\;.
\end{equation}
By substituting these into (\ref{sufficient solution}), and using
(\ref{X}), one can read off
\begin{equation}
\mathcal{P}_i( \underline{\mu}_{i}, \underline{m}_i -
\underline{\mu}_i ) = v( \underline{\mu}_{i}) \,
w_i(\underline{m_i}-\underline{\mu}),
\end{equation}
then from the definition of $\mathcal{P}$ given in (\ref{P}), equation
(\ref{mass transfer probs}) immediately follows.

\section{Continuous time limit}
\label{continuous time}
The discrete time update contains as a special case random sequential
update which in turn contains the continuous time limit. The random
sequential limit emerges when the mass transport probabilities are
proportional to a small time step $dt$ such that at most one mass
transfer event on the lattice takes place within a time step. Then the
limit $dt\to 0$ yields continuous time dynamics where the masses are
transferred with a rate per unit time. We take the random sequential
limit first by redefining
\begin{equation}
v(\underline{\mu}) = \delta^Q(\underline{\mu}) +
x(\underline{\mu}) {\rm d}t\;.
\end{equation}
Upon substitution into (\ref{single-site weights}) this yields
\begin{equation}
f_i(\underline{m}_i) = w_{i}(\underline{m}_i)
\delta^Q(\underline{\mu}) + {\rm d}t [x*w_i] (\underline{m}_i) +
\mathcal{O}({\rm d}t^2)\;,
\end{equation}
for the single-site weights, and from (\ref{mass transfer probs}) it
yields 
\begin{eqnarray}
\psi_i( \underline{\mu} | \underline{m}_i ) &=&
\delta^Q(\underline{\mu}) - \frac{{\rm
d}t}{w_{i}(\underline{m}_i)} \delta^Q(\underline{\mu}) [x *
w_{i}](\underline{m}_i) 
\nonumber \\ && 
+ {\rm d}t
\frac{x(\underline{\mu}) w_{i}(\underline{m}_i -
\underline{\mu})}{w_{i}(\underline{m}_i)} 
+
\mathcal{O}({\rm d}t^2)\;,
\end{eqnarray}
for the mass transfer probabilities.  In the next step, we take the
continuous time limit by taking $dt \to 0$, hence the single-site
weights are
\begin{equation} \label{continuous time f} 
f_i(\underline{m}) = w_{i}(\underline{m})\;,
\end{equation}
and the hop rates are read off as
\begin{equation} \label{condition}
u_{i}(\underline{\mu} | \underline{m}_i) =
\frac{x(\underline{\mu})\;w_{i}(\underline{m} -
  \underline{\mu})}{w_{i}(\underline{m})} \;.
\end{equation}
This hop rate represents the rate at which masses $\mu^{(1)}, \ldots,
\mu^{(Q)}$ are simultaneously transferred from site $i$ to site $i+1$.
If the hop rates can be written in the form (\ref{condition}), the
steady state factorises with single-site weights given by
(\ref{continuous time f}).

We note that the form (\ref{condition}) implies a constraint on the
choice of hop rates. Viewed as a recursion, one can express
$w_{i}(\underline{m})$ in terms either $w_{i}(\underline{m} -
\underline{\mu})$ or $w_{i}(\underline{m} - \underline{\mu}')$, before
a second application of the recursion yields an expression in terms of
$w_{i}(\underline{m} - \underline{\mu}-\underline{\mu}')$; this final
expression must be the same regardless of the intermediate step, which
demands that the hop rates satisfy
\begin{equation} \label{hop rate constraint}
\frac{u_{i}(\underline{\mu} | \underline{m})}
{u_{i}(\underline{\mu} | \underline{m} - \underline{\mu}')} =
\frac{u_{i}(\underline{\mu}' | \underline{m})}
{u_{i}(\underline{\mu}' | \underline{m} - \underline{\mu})}\;,
\end{equation}
for all $i, \underline{m}, \underline{\mu}$ and $\underline{\mu}'$.
This constraint is not an additional constraint to (\ref{condition}):
if the hop rates can be written in the form (\ref{condition}), they
automatically satisfy (\ref{hop rate constraint}). However, it
provides an alternative test for factorisation for a given set of hop
rates. Moreover, this constraint represents a necessary and sufficient
condition for factorisation.

\subsection{Two-species ZRP with disorder} \label{disordered 2ZRP}
The ZRP with two species of particles is a discrete-mass model in
which single units of mass, either of species 1 or of species 2, are
transferred to the nearest neighbour site to the right \cite{GS03,
EH03}. Here, the model is generalised to include disorder in the hop
rates. In the current notation, hop rates for species 1 and species 2
are, respectively,
\begin{eqnarray} \label{d2zrp1}
u_{i}( 1, 0 | m_i^{(1)}, m_i^{(2)}) =
\frac{x(1,0) \; f_{i}(m_i^{(1)} - 1, m_i^{(2)})}{f_{i}(m_i^{(1)},
  m_i^{(2)})} \;, \\ \label{d2zrp2}
u_{i}( 0, 1 | m_i^{(1)}, m_i^{(2)}) =
\frac{x(0,1) \; f_{i}(m_i^{(1)}, m_i^{(2)}-1)}{f_{i}(m_i^{(1)},
  m_i^{(2)})} \;,
\end{eqnarray}
which imply the constraint
\begin{equation} \label{d2zrpcon}
\frac{u_{i}( 1, 0 | m_i^{(1)}, m_i^{(2)})}{u_{i}( 1, 0 |
  m_i^{(1)}, m_i^{(2)} - 1)} = \frac{u_{i}( 0, 1 | m_i^{(1)},
  m_i^{(2)})}{u_{i}( 0, 1 | m_i^{(1)} - 1, m_i^{(2)})}\;,
\end{equation}
The single-site weights, obtained by iterating (\ref{d2zrp1},\ref{d2zrp2}),
are
\begin{equation} \label{d2zrpf}
f_{i}(m^{(1)}, m^{(2)}) = \prod_{x=1}^{m^{(1)}} u_{i}( 1, 0 |
x, m^{(2)})^{-1} \prod_{y=1}^{m^{(2)}} u_{i}( 0, 1 | 0,
y)^{-1}\;,
\end{equation}
having set $f_i(0,0)=1$ and absorbed $x(1,0)$ and $x(0,1)$ into the
definitions of the hop rates without loss of generality. So, for any
given hop rates, (\ref{d2zrpcon}) represents a necessary and
sufficient condition for the steady state to factorise with
$f_{i}(\underline{m})$ given by (\ref{d2zrpf}). This generalises the
results of \cite{GS03, EH03} to the disordered case.
 
\subsection{A two-species model}
\label{two species model}
In this model, again we consider two species of discrete masses. Now,
the number of particles of a species which hops depends on the number
of particles of the other species at the departure site. Hence, we
consider the following rates: if $m_i^{(1)} > m_i^{(2)} > 0$, then
$m_i^{(2)}$ particles of species 1 hop with a rate
\begin{equation}
u(m_i^{(2)}, 0 | m_i^{(1)}, m_i^{(2)}) = \frac{f(m_i^{(1)} -
  m_i^{(2)}, m_i^{(2)})}{f(m_i^{(1)}, m_i^{(2)})}\;;
\end{equation} 
if $m_i^{(2)} > m_i^{(1)} > 0$, then $m_i^{(1)}$ particles of species 2
hop with a rate
\begin{equation}
u(0, m_i^{(1)} | m_i^{(1)}, m_i^{(2)}) = \frac{f(m_i^{(1)}, m_i^{(2)} -
  m_i^{(1)})}{f(m_i^{(1)}, m_i^{(2)})}\;;
\end{equation}
if $m_i^{(1)} = m_i^{(2)} = m > 0$ then, with equal probability,
either $m$ particles of species 1 hop with a rate
\begin{equation}
u(m, 0 | m, m) = \frac{f(0,m)}{f(m , m)}\;,
\end{equation}  
or $m$ particles of species 2 hop with a rate
\begin{equation}
u(0, m | m, m) = \frac{f(m,0)}{f(m , m)}\;.
\end{equation}
If only a single species is present at a site, single particles
hop with rates
\begin{equation}
u(1,0|m,0) = \frac{f(m-1,0)}{f(m,0)}\;,\qquad u(0,1|0,n) =
\frac{f(0,n-1)}{f(0,n)}\;.
\end{equation} 

A simple choice for the rates is $u(n,0|m,n) = (1+n/m)^b$ for $m \geq
n > 0$ and $u(0,m|m,n) = (1+m/n)^b$ for $n\geq m>0$; the single-site
weights for these rates are given by $f(m,n) = (m+n)^{-b}$. Though the
model presented here is new, single-site weights of this form have
been considered in \cite{AEH} where it was shown that a condensation
transition occurs above a critical particle density for
$b>3$. Ordinarily, the two species condense at completely independent
sites; in the present model however, an exclusion interaction exists
between the condensates: by considering the hop rates in the limit of
large $n$ and $m$, it is clear that if both species condense at the
same time, they do so at randomly located sites subject to the
condition that the condensates of each species do not occupy the same
site.
 
\section{Generalisation to hypercubic and arbitrary lattices}
\label{generalisation}
The models considered above can be generalised to $Q$-species
mass-transfer models on arbitrary lattices. The derivation of the
conditions for factorisation can also be generalised along the lines
of \cite{EMZ06}. Though this derivation is straightforward, its simplicity
is rather obscured under complicated notation; therefore we present in
this section the results without providing details. We also provide
concrete examples of models with factorised steady states in order to
illustrate the utility of the results.

\subsection{Generalised model}
Again, we begin with discrete time dynamics whereby each site is
updated in parallel. The new ingredient is that $Q$ fixed sets of
directed links now connect certain pairs of sites --- each set is
associated with a particular species and mass of any species can only
be transferred along associated directed links.

To define the update rules, we define a mass transfer matrix for each
species $q$ with elements $\mu_{ij}^{(q)}$. These elements represent
the mass of species $q$ which is transferred from site $i$ to site $j$
during a time step. The vector $\underline{\mu}_{ij} =
(\mu_{ij}^{(1)}, \ldots, \mu_{ij}^{(Q)})$ represents the masses of all
species which move from $i$ to $j$; if there is no directed link for
species $q$ pointing from $i$ to $j$, then $\mu_{ij}^{(q)}=0$
identically. Also, $\mu_{ii}^{(q)}$ is in general nonzero and
represents the mass of species $q$ at site $i$ that does not hop
during the time-step.

The elements $\mu_{ij}^{(q)}$ which are not set to zero for all times
(i.e., those which are accompanied by a directed link pointing from
$i$ to $j$) are determined by the mass transfer probabilities $\psi_i(
\{ \underline{\mu}_{ij} \} | \underline{m}_i )$. The set $\{
\underline{\mu}_{ij} \}$ contains all sites $j$ to which $i$ is
connected by a directed link for at least one species and includes
site $i$ itself. Thus it specifies the mass transferred for each and
every species which can hop from $i$ to $j$.

As before, the dynamics conserve the total mass of each species in the
system. The sum over the elements in a row of the mass transfer matrix
for species $q$ is equal to the mass of species $q$ at the site before
the update: $\sum_j \mu_{ij}^{(q)} = m_i^{(q)}$. Similarly, the sum
over column elements represents the mass $m_i^{\prime (q)}$ of species
$q$ at $i$ after the update: $\sum_j \mu_{ji}^{(q)} = m_i^{\prime
(q)}$.

\subsection{Conditions under which the steady state factorises} \label{gen ss}
Using the multi-species procedure presented in section
\ref{derivation} and following the technique outlined in \cite{EMZ06}
for one species of mass on an arbitrary lattice, one can establish
conditions on the generalised mass transfer probabilities under which
the steady state assumes the factorised form (\ref{fss}).

We find a sufficient condition for factorisation (but no longer
necessary in general) is that one can find functions
$v_{ij}(\underline{\mu}_{ij})$ such that the mass transfer
probabilities can be expressed in the form
\begin{equation} \label{gen mass transfer probs}
\psi_i( \{ \underline{\mu}_{ij} \} | \underline{m}_i ) = 
\frac{\prod_j v_{ij}(\underline{\mu}_{ij})}{\left[\prod_{*j}
    v_{ji}\right](\underline{m}_{i})}\;,
\end{equation}
subject to the constraint $\sum_j \mu_{ij}^{(q)} = m_i^{(q)}$ for all
species $q$. Here, the product in the numerator is over
sites $j$ which have a directed link coming from site $i$ for at least
one species; the product in the denominator is over all sites which have
at least one
directed link pointing into $i$ and the $*$ denotes a convolution
form. If the mass transfer probabilities are of
the form (\ref{gen mass transfer probs}), then the steady state factorises
with single-site weights $f_i( \underline{m}_i )$ given by
\begin{equation} \label{gen single-site weights}
f_i( \underline{m}_i ) = \left[\prod_{*j} 
  v_{ji}\right](\underline{m}_{i})\;,
\end{equation}
which again is a multiple convolution over those sites which have a
directed link into $i$ for at least one species.  

A second condition for factorisation is that the functions
$v_{ij}(\underline{\mu}_{ij})$ satisfy
\begin{equation} \label{v constraint}
\left[\prod_{*j} v_{ij}\right](\underline{m}_{i}) = \left[\prod_{*j}
  v_{ji}\right](\underline{m}_{i})\;,
\end{equation}
which emerges as a constraint on the geometry of the underlying
lattice. This constraint will typically limit the choices allowed for
the dynamics on inhomogeneous lattices. On homogeneous lattices, there
exist two particular classes of dynamics for which the constraint is
always satisfied: (i) detailed balance --- in cases where the dynamics
are symmetric, i.e., $v_{ij}(\underline{\mu}_{ij}) =
v_{ji}(\underline{\mu}_{ji})$ the constraint (\ref{v constraint}) is
always satisfied; (ii) pair-wise detailed balance --- the constraint
(\ref{v constraint}) is always satisfied in cases where, for each
species, there exists a site $i$ with a link into $j$ for every link
from $j$ to another site $k$, such that $v_{ij}(\underline{\mu}_{ij})
= v_{jk}(\underline{\mu}_{jk})$. A simple example is asymmetric
nearest neighbour mass transport in one dimension, where transfer to $i$
from $i-1$ can be paired with transfer from $i$ to $i+1$. There are many
more ways to satisfy the constraint, some of which are discussed
below, but dynamics which fall into either of the above two categories
are sufficiently prevalent in mass transport models that we wish to
emphasise that the constraint is automatically fulfilled in these two
cases.

The general conditions under which (\ref{gen mass transfer probs})
represents a necessary condition seem difficult to formulate
\cite{EMZ06}: though necessary for totally asymmetric mass transport
in one dimension, as outlined above, it is not necessary for symmetric
mass transport. It is also necessary for mass transfer on the complete
graph.

\subsection{Continuous time limit}
The random sequential then continuous time limits are obtained as
before, by redefining
\begin{equation}
v_{ij}(\underline{\mu}_{ij}) = \delta^Q(\underline{\mu}_{ij}) +
x_{ij}(\underline{\mu}_{ij}) {\rm d}t\;.
\end{equation}
In the limit $dt \to 0$, the single-site weights assume the form
\begin{equation} \label{gen continuous time f} 
f_i(\underline{m}_i) = v_{ii}(\underline{m}_i)\;,
\end{equation}
and hop rates are obtained as 
\begin{equation} \label{gen condition}
u_{ij}(\underline{\mu} | \underline{m}_i) =
\frac{x_{ij}(\underline{\mu})\;v_{ii}(\underline{m}_i -
  \underline{\mu})}{v_{ii}(\underline{m_i})} \;.
\end{equation}
This hop rate represents the rate at which masses $\mu^{(1)}, \ldots,
\mu^{(Q)}$ are simultaneously transferred from site $i$ to site $j$.
Finally, in the continuous time limit, the constraint acquires the
form
\begin{equation} 
\sum_{j\neq i} [x_{ij}*v_{ii}](\underline{m}_i) = \sum_{j\neq i}
    [x_{ji}*v_{ii}](\underline{m}_i)\;.
\end{equation}
By taking the Laplace transform, cancelling common factors then inverting
Laplace transform back, the constraint can be written
\begin{equation} \label{constraint}
\sum_{j\neq i} x_{ij}(\underline{\mu}) = \sum_{j\neq i}
x_{ji}(\underline{\mu})\;.
\end{equation}
Therefore, if the hop rates can be written in the form (\ref{gen
condition}), and provided the constraint (\ref{constraint}) is
satisfied, the steady state factorises with single-site weights given
by (\ref{gen continuous time f}).

\subsection{Generalised $Q$-species mass-transfer model in one
  dimension}
\label{1d gen Q-ZRP}
In the case where each particle species is transferred between nearest
neighbour sites on the same one-dimensional lattice, the constraint
(\ref{constraint}) implies
\begin{equation} \label{1d walker}
x_{i, i{+}1}(\underline{\mu}) + x_{i, i{-}1}(\underline{\mu}) = 
x_{i{+}1, i}(\underline{\mu}) + x_{i{-}1, i}(\underline{\mu})\;,
\end{equation}
and the general solution has been obtained in \cite{D83}; this
solution must be used in order to allow disorder in the
asymmetry. Here, we take the pair-wise balance solution
\begin{eqnarray} \label{sol}
x_{i, i{+}1}(\underline{\mu})  = p(\underline{\mu})\;, \\
x_{i{+}1, i}(\underline{\mu})  = q(\underline{\mu})\;, 
\end{eqnarray}
for all $i$. Hence the steady state factorises if the hop rates can be
written in the form
\begin{eqnarray} \label{form}
u_{i, i{+}1}( \underline{\mu} | \underline{m}_i) =
\frac{p(\underline{\mu})\;f_{i}(\underline{m}_i -
  \underline{\mu})}{f_{i}(\underline{m_i})} \;, \\
u_{i, i{-}1}( \underline{\mu} | \underline{m}_i) =
\frac{q(\underline{\mu})\;f_{i}(\underline{m}_i -
  \underline{\mu})}{f_{i}(\underline{m_i})} \;,
\end{eqnarray}
for all $i$. Note that some disorder is still incorporated through the
site dependence in $v_{ii}(\underline{m})$: the disorder is in the
departure rates rather than the asymmetry.

The two-species, totally asymmetric limit $q(\underline{\mu})=0$ with
transfer of single units of discrete mass recovers the limit discussed
in section \ref{disordered 2ZRP}.

\subsection{Generalised Q-species mass transfer on a hypercubic lattice}
Here, we consider homogeneous, symmetric dynamics on a hypercubic
lattice: masses $\underline{\mu} = (\mu^{(1)}, \ldots, \mu^{(Q)})$ are
transferred to a randomly selected nearest neighbour site with a rate
$u( \underline{\mu} | \underline{m})$.

In this case, the dynamics satisfy detailed balance so the constraint
(\ref{constraint}) is easily seen to be satisfied for
$x_{ij}(\underline{\mu})$ independent of $i$ and $j$. The sufficient
condition for factorisation then is that the hop rates satisfy the
constraint
\begin{equation}
\frac{u(\underline{\mu} | \underline{m}_i)}
{u(\underline{\mu} | \underline{m}_i - \underline{\mu}')} =
\frac{u(\underline{\mu}' | \underline{m}_i)}
{u(\underline{\mu}' | \underline{m}_i - \underline{\mu})}\;,
\end{equation}
in which case the single-site weights are determined from
\begin{equation}
u( \underline{\mu} | \underline{m}_i) = \frac{f(\underline{m}_i -
  \underline{\mu})}{f(\underline{m_i})} \;,
\end{equation}
where the constant factors $x(\underline{\mu})$ have been absorbed
into $u$. Thus, since the rates are homogeneous, the single-site
weights are independent of $i$. This generalises the results of
\cite{GL06} to an arbitrary number of species of
particles. Generalisations to include disorder or to partial asymmetry
along the lines of the section \ref{1d gen Q-ZRP} are straightforward.

\subsection{Two-species ZRP on an arbitrary lattice}
In this section, we consider the two-species ZRP again, but this time
on an arbitrary lattice. We consider the case where both species hop
on the same underlying lattice, with hop rates from site $i$ to site
$j$ given by
\begin{equation} \label{arb rate 1}
u_{ij}( 1,0 | m_i^{(1)}, m_i^{(2)}) = y_{ij} \; \alpha_i^{(1)}(m_i^{(1)},
m_i^{(2)})\;,
\end{equation}
\begin{equation} \label{arb rate 2}
u_{ij}( 0,1 | m_i^{(1)}, m_i^{(2)}) = y_{ij} \; \alpha_i^{(2)}(m_i^{(1)},
m_i^{(2)})\;,
\end{equation}
for species 1 and 2, respectively. Here, $\alpha_i^{(q)}(m_i^{(1)},
m_i^{(2)})$ represents the total departure rate of particles of
species $q$ from site $i$, for $q=1,2$; $y_{ij}$ represents the
probability that a particle is transferred from site $i$ to site
$j$. The same probabilities are used for both species because they are
both moving on the same underlying lattice (though this is
straightforward to generalise).

The choices 
\begin{equation}
x_{ij}(\underline{\mu}) = y_{ij} p_i \;,
\end{equation}
for $\underline{\mu} = (1,0)$ or $(0,1)$, and
\begin{equation} \label{arb2zrp1}
\alpha_i^{(1)}(m_i^{(1)}, m_i^{(2)})=\frac{p_i f_{i}(m^{(1)}{-}1,
  m^{(2)})}{f_{i}(m^{(1)}, m^{(2)})}\;, 
\end{equation}
\begin{equation} \label{arb2zrp2}
  \alpha_i^{(2)}(m_i^{(1)}, m_i^{(2)})=\frac{p_i
  f_{i}(m^{(1)}, m^{(2)}{-}1)}{f_{i}(m^{(1)}, m^{(2)})}\;,
\end{equation}
are of the form (\ref{gen condition}) from which the single-site weights
are obtained as
\begin{equation} \label{arb f}
f_{i}(m^{(1)}, m^{(2)}) = p_i^{m_i^{(1)} + m_i^{(2)}}
\prod_{x=1}^{m^{(1)}} \alpha_{i}^{(1)}( x, m^{(2)})^{-1}
\prod_{y=1}^{m^{(2)}} \alpha_{i}^{(2)}( 0, y)^{-1}\;.
\end{equation}
From the constraint (\ref{constraint}), the $p_i$'s are determined by
\begin{equation} \label{p sol}
p_i = \sum_{j\neq i} y_{ji} p_j\;,
\end{equation}
which is just the solution for the steady state of a single random
walker moving on the lattice with hopping probabilities defined by the
$y_{ij}$ (c.f. (\ref{1d walker})).

Again the two equations (\ref{arb2zrp1}, \ref{arb2zrp2}) imply a constraint
on the possible choices of hop rates for which the steady state still
factorises: 
\begin{equation}
\frac{\alpha_{i}^{(1)}( m_i^{(1)}, m_i^{(2)})}{\alpha_{i}^{(2)}(
  m_i^{(1)}, m_i^{(2)} - 1)} = \frac{\alpha_{i}^{(2)}( m_i^{(1)},
  m_i^{(2)})}{\alpha_{i}^{(2)}( m_i^{(1)} - 1, m_i^{(2)})}\;.
\end{equation}
If this equation is satisfied, a model with rates (\ref{arb rate 1},
\ref{arb rate 2}) factorises with single-site weights given by
(\ref{arb f}) with the $p_i$'s determined by (\ref{p sol}) --- thus it
is the $p_i$'s which contain the information about the structure of
the underlying lattice.
 
\section{Conclusion}
We have found conditions for factorised steady states applicable to a
very wide class of models. The results have been exploited to write
down the exact steady states for several models which generalise
previously solved cases. The interplay between conservation laws,
geometry, disorder and the nature of interactions can all be explored
within the framework of the steady states derived here and merit
further investigation.

As intimated in the introduction, condensation is of particular
interest in mass transfer models. For homogeneous two-species systems,
the general conditions for condensation are well understood at the
level of the steady state \cite{G}. However the models considered
here are substantially more general, therefore it is of interest to
establish whether new condensed phases emerge. Even for the two
species model considered in section \ref{two species model}, which has
a steady state in the class considered in \cite{G}, the exclusion
interaction between condensates seems to be a property of the dynamics
which has not been observed in previously studied models.

Moreover, the coarsening dynamics in two-species models have been
studied using heuristic, quasi-static scaling arguments \cite{GH05},
and are far richer than their single-species counterparts. It is
reasonable to suppose that the wider class of models solved here will
further enrich the phenomenology of the coarsening dynamics.

Finally, the question of the structure of steady states beyond the
factorised form remains open \cite{EHM06} but must be addressed in
order to understand behaviour observed in non-factorisable models
\cite{G06}.

\ack The author thanks Martin Evans for useful discussions and
acknowledges the SUPA and the EPSRC for support under programme grant
GR/S10377/01.


\section*{References}
\begin{thebibliography}{00}

\bibitem{CSS00} 
Chowdhury D, Santen L and Schadschneider A, 2000 Physics Reports {\bf
  329} 199 

\bibitem{OEC98}  
O'Loan OJ, Evans MR and Cates ME, 1998 Phys. Rev. E {\bf 58} 1404

\bibitem{MWL02} 
van der Meer D, van der Weele K and Lohse D, 2002 Phys. Rev. Lett. {\bf
  88} 174302 

\bibitem{J05}
Jain K, 2005 Phys. Rev. E {\bf 72} 017105
 
\bibitem{MKB98}
Majumdar SN, Krishnamurthy S and Barma M, 1998 Phys. Rev. Lett.
{\bf 81} 3691 

\bibitem{KLMST02}
Kafri Y, Levine E, Mukamel D, Sch{\"u}tz GM and
T{\"o}r{\"o}k J, 2002 Phys. Rev. Lett. {\bf 89} 035702 

\bibitem{S70}
Spitzer F, 1970 Adv. Math. {\bf 5} 246

\bibitem{EH05}
Evans MR and Hanney T, 2005 J. Phys. A {\bf 38} R195
 
\bibitem{KG00}
Krug J and Garcia J, 2000 J. Stat. Phys. {\bf 99} 31 

\bibitem{RM00}
Rajesh R and Majumdar SN, 2000 J. Stat. Phys. {\bf 99} 943 

\bibitem{EMZ06}
Evans MR, Majumdar SN and Zia RKP, 2006 J. Phys. A {\bf 39} 4859

\bibitem{schutz review}
Sch{\"u}tz GM, 2003 J. Phys. A {\bf 36} R339
 
\bibitem{HE04}
Hanney T and Evans MR, 2004 Phys. Rev. E {\bf 69} 016107 

\bibitem{MMWL02}
Mikkelsen R, van der Meer D, van der Weele K and Lohse D, 2002
Phys. Rev. Lett. {\bf 89} 214301

\bibitem{DMS03}
Dorogovtsev SN, Mendes JFF and Samukhin AN, 2003
Nucl. Phys. B {\bf 666} 396
 
\bibitem{AEH}
Angel AG, Hanney T, and Evans MR, 2006 Phys. Rev. E {\bf 73} 016105

\bibitem{EMZ04}
Evans MR, Majumdar SN and Zia RKP, 2004 J. Phys. A {\bf 37} L275

\bibitem{EMZ04a}
Zia RKP, Evans MR and Majumdar SN, 2004
J. Stat. Mech.:Theor. Exp. L10001 

\bibitem{GS03}
Gro{\ss}kinsky S and Spohn H, 2003 Bull. Braz. Math. Soc. {\bf 34} 489

\bibitem{EH03}
Evans MR and Hanney T, 2003 J. Phys. A {\bf 36} L441

\bibitem{D83}
Derrida B, 1983 J. Stat. Phys. {\bf 31} 433

\bibitem{GL06}
Greenblatt RL and Lebowitz JL, 2006 J. Phys. A {\bf 39} 1565

\bibitem{G}
Gro{\ss}kinsky S, {\it to be published}

\bibitem{GH05}
Gro{\ss}kinsky S and Hanney T, 2005 Phys. Rev. E {\bf 72} 016129

\bibitem{EHM06}
Evans MR, Hanney T and Majumdar SN, 2006 Phys. Rev. Lett. {\bf
  97} 010602
 
\bibitem{G06}
Godr\`eche C, 2006 J. Phys. A {\bf 39} 9055

\end{thebibliography}
\end{document}